\documentclass{b98proc}
\usepackage{amsmath,pennames}

\def\SELEX{\textsc{Selex} }
\def\r2{\langle r^2 \rangle}
\def\Q2{$Q^2$}

\def\gevc{GeV/$c$ }
\def\gev2c2{GeV$^2$/$c^2$}
\def\fm2{\text{fm}^2}

\def\Journal#1#2#3#4{{#1} {\bf #2}, #3 (#4)}

\def\NPA{{\em Nucl. Phys.} A}
\def\NPB{{\em Nucl. Phys.} B}

\def\PLB{{\em Phys. Lett.} B}

\def\PRL{\em Phys. Rev. Lett.}

\def\PRD{{\em Phys. Rev.} D}
\def\PRC{{\em Phys. Rev.} C}

\begin{document}

\title{MEASUREMENT OF THE $\Sigma^-$ CHARGE RADIUS AT SELEX}

\author{I. ESCHRICH}
\address{Max-Planck-Institut f\"ur Kernphysik, Heidelberg, Germany}
\author{on behalf of the SELEX Collaboration\footnotemark}

\maketitle

\abstracts{
The charge radii of \PgSm\ and \Pgpm\ have been determined
by direct elastic scattering on shell electrons.
The measurement was performed in the framework of the
\SELEX (E781)
charm hadroproduction experiment at Fermilab which employs a 600~\gevc
high-intensity \PgSm/\Pgpm\ beam and a three-stage magnetic
spectrometer covering $0.1 \le x_F \le 1.0 $.
Scattering angles and momenta of both
hadron and electron were measured with high precision
using silicon microstrip detectors, thus
allowing for a segmented solid target.
Two TRDs provided full particle identification.
A preliminary result for the \PgSm\ charge radius
for a four-momentum transfer squared of
$0.03~\text{GeV}^2/c^2 \le Q^2 \le 0.16~\text{GeV}^2/c^2$ will be reported.
In a parallel analysis the \Pgpm\ charge radius has been determined for
$0.03~\text{GeV}^2/c^2 \le Q^2 \le 0.2~\text{GeV}^2/c^2$, and is found to be
consistent with previous experiments.
}

\def\thefootnote{*}
\footnotetext{
  Ball State University,
  Bogazici University Istanbul,
  Carnegie Mellon University,
  Centro Brasileiro de Pesquisas Fisicas Rio de Janeiro,
  Fermilab,
  IHEP Bejing,
  IHEP Protvino,
  ITEP Moscow,
  Max-Planck-Institut f\"ur Kernphysik Heidelberg,
  Moscow State University,
  Petersburg Nuclear Physics Institute,
  Tel Aviv University,
  Universidad Autonoma de San Luis Potosi,
  Universidade Federal da Paraiba,
  University of Bristol,
  University of Hawaii,
  University of Iowa,
  University of Michigan-Flint,
  University of Rochester,
  University of Rome 'La Sapienza' and INFN Rome,
  University of S\~ao Paulo,
  University of Trieste and INFN Trieste.
  }
  
\section{Introduction}
\label{sec:intro}

We report here on first results of a measurement of the
\PgSm\ and \Pgpm\ charge radii by direct elastic scattering off atomic shell
electrons.

Hadrons as we understand them today are composite systems which we
characterize by their static properties.
One static property which reflects the
phenomenon unique to hadrons -- quark confinement -- is the size
of the particle.

The definition of the radius of a hadron depends on the probe used to
measure it. A strong-interaction mean squared radius can be extracted
from hadron-proton
collisions. The hadron-electron interaction, on the other hand, yields
the mean squared charge radius:
Elastic scattering of an electron off a charged hadron is modified from
a point interaction by the form factor $F(Q^2)$ where \Q2 is the
four-momentum transfer squared.
At zero momentum transfer the mean squared charge radius is related to
the slope of the form factor by
\begin{equation*}
  \label{equ:rad_from_ff}
  \r2 = -6\hbar^2 \frac{dF(Q^2)}{dQ^2}\Bigg|_{Q^2=0}.
\end{equation*}
Unfortunately, charge radii are known only for five different
hadrons so far.
The fact that
the \PKm\ radius has been found to be smaller than that of the \Pgpm\
by $\sim 0.1\: \fm2$
suggests that the size of a hadron is related to the flavor composition
of its constituent quarks. There is supporting evidence from a 
study of strong interaction radii \cite{povh90} which finds that
replacing an \textit{up} or \textit{down} quark by a
\textit{strange} quark in a baryon decreases its radius by approximately
0.08~fm$^2$.
Consequently the \PgSm\ radius should be smaller than
the proton radius, and larger than the \PgXm.
The definition of a strong-interaction radius, however, is model-dependent.
The significance of the above observation is therefore limited unless
validated by a systematic study of hyperon charge radii.

\section{Experimental setup}
\label{sec:exp}

The 800~\gevc proton beam from the Fermilab Tevatron was used to produce
a beam consisting to approximately equal parts of \PgSm\ and \Pgpm\ at
600~\gevc with a momentum spread of $\pm$8~\%. 
\begin{figure}[h]
  \begin{center}
    \leavevmode
    \epsfig{file=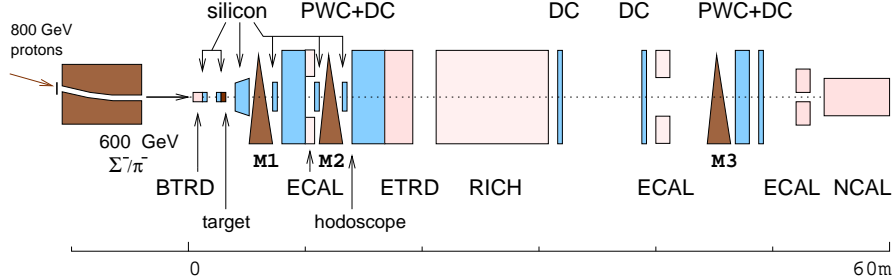, width=\textwidth}
  \end{center}
  \caption{Schematic layout of the \SELEX spectrometer. Three analyzing
    magnets ({\tt M1/M2/M3}, indicated by prisms) define sub-spectrometers
    dedicated to different momentum regions.
    Transverse dimensions are not to scale. }
  \label{fig:selex}
\end{figure}
The primary objective of the experiment being the hadroproduction and
spectroscopy of charm baryons in the forward hemisphere, \SELEX was laid
out as a 3-stage magnetic spectrometer as shown in Fig.~\ref{fig:selex}.
Beam particles were identified by a transition radiation detector (BTRD).
Interactions took place in a target stack of two Cu and three C foils
adding up to 5~\% of an interaction length for protons. Downstream of the
targets 20 silicon planes of 20-25~$\mu$m strip pitch provided good vertex
resolution.

The M1 and M2 magnets implemented momentum deflections of 2.5~\gevc
and 15~\gevc, respectively. Each stage of the spectrometer was equipped
with proportional and drift chambers for tracking and a lead glass
calorimeter (ECAL). In addition, 50~$\mu$m pitch silicon detectors were used
close to the beam axis downstream of the M1 and M2 magnets to improve
the resolution for high momenta. A second transition radiation detector
(ETRD) provided electron identification.
\SELEX was also equipped with a ring-imaging \v Cerenkov counter (RICH)
for separation of \Pp, \PK, and \Pgp. A third spectrometer stage
aided in the reconstruction of large-momentum \PgL.
A hadronic calorimeter (NCAL) concluded the setup.


For hadron-electron elastic scattering,
two hits in the negative and none in the positive half of a hodoscope
downstream of the second magnet in coincidence with a multiplicity of
two in a set of scintillation counters 3~cm downstream of the target
constituted a valid trigger condition.
The typical trigger rate at this level was 3000 per 20-second spill at
a beam rate of 10$^7$ particles per spill.
An online filter
performed a preliminary
track reconstruction in the M2 spectrometer. Requiring at least one
track with negative and none with positive slope together with other
conditions crucial to a complete reconstruction reduced this sample
by a factor of 1:1.7


\section{Data Analysis}
\label{sec:ana}

In the 1997 run \SELEX has recorded 215~million candidates for hadron-electron
scattering with \PgSm/\Pgpm-beam.
In preparation for a first analysis with the software tools
available at that time the negative-beam sample was stripped to 10~\% of
its original size by cutting on an electron signature in the ETRD, unambiguous
identification of the beam particle by the BTRD, and a two-negative-track
event topology in the M2 spectrometer. A second-stage strip required a
two-prong vertex, again reducing the sample by a factor of 10.

Out of the stripped data sample described above, 12,000 \PgSm-~electron and
26,000 \Pgpm- ~electron elastic scattering events were extracted.
For each event, the incoming and outgoing tracks in the vertex
were required to be coplanar. Particle identification for the two outgoing
tracks was performed by combining ETRD information with kinematic
constraints. Events with ambiguous particle identification were discarded.
\begin{figure}[h]
  \begin{center}
    \leavevmode
    \epsfig{file=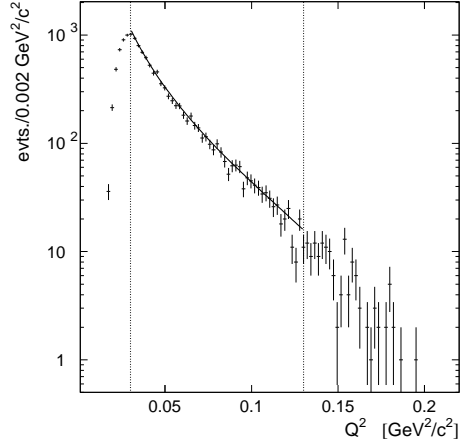, height=6.5cm}
  \end{center}
  \caption{\Q2 distribution of \PgSm-electron scattering events.
    Vertical lines indicate the region accepted for fitting. }
  \label{fig:q2dist}
\end{figure}
For \PgSm, decays upstream of
the M2 chambers were rejected by requiring the scattered beam particle to
have at least 60~\% of the incoming beam particle's momentum.
Finally, electron momentum and scattering angle had to match their expected
kinematic relation to better than 10~\%. 

The charge radii were determined by fitting the differential cross section
with an assumed radius as single parameter to the observed distribution of
the four-momentum transfer squared \Q2 (Fig.~\ref{fig:q2dist}).
Since the shape of the \Q2-distribution yields the radius no absolute
normaliziation is needed.
In this first analysis, \Q2 was calculated from the beam momentum and
the scattering angle of the electron. From Monte Carlo studies the
\Q2 resolution was estimated to be 1.5~\%. 
Preliminary acceptance studies were performed using generated elastic
scattering events
embedded in real data. The geometrical and reconstruction-dependent
acceptance was modeled and a preliminary evaluation of the trigger
efficiency performed.

\begin{figure}[h]
  \begin{center}
    \leavevmode
    \epsfig{file=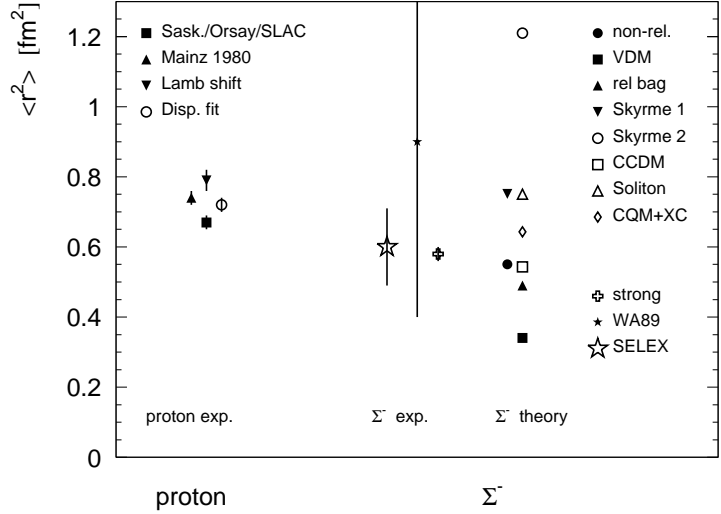, height=8.25cm}
  \end{center}
  \caption{The \PgSm\ charge radius compared to
    various results for the proton radius (left):
    \textit{Sask./Orsay/SLAC}\protect\cite{hand63,murphy74},
    \textit{Mainz}\protect\cite{simon80a},
    \textit{dispersion-theoretical fit} to all of
    above\protect\cite{mergell96}, and
    \textit{Lamb shift}\protect\cite{udem97}.
     -- Experimental results (center):
    \textit{SELEX}: this measurement,
    \textit{WA89}: WA89 result\protect\cite{wa89_radius_paper},
    \textit{strong}: strong interaction radius\protect\cite{povh90}. -- 
    The predictions for \PgSm\ refer to the following models:
    \textit{non-rel.}: non-relativistic
    quark model, \textit{VDM}: vector dominance model,
    \textit{rel bag}: relativistic bag model (all three values from
    \protect\cite{povh90}),
    \textit{Skyrme 1}: Skyrme model\protect\cite{kunz91},
    \textit{Skyrme 2}: Skyrme model\protect\cite{park92},
    \textit{CCDM}: Chiral color dielectric model\protect\cite{sahu94},
    \textit{Soliton}: Soliton model\protect\cite{kim96},
    \textit{CQM+XC}: Chiral constituent quark model including
    exchange currents\protect\cite{wagner98b}.}
  \label{fig:sigmaresult}
\end{figure}
For the \PgSm\ data, a \Q2 region with flat acceptance was chosen for
fitting the radius. For the \Pgpm\ data, an acceptance correction was applied.
Each event was normalized to its individual beam momentum to eliminate
effects of the beam momentum spread. An unbinned maximum likelihood
fit using dipole electric and magnetic form factors for the \PgSm\
yields a mean squared charge radius of
\begin{equation*}
  \label{equ:final_result}
  \r2_{\Sigma^-} = 0.60 \pm 0.08 \:(stat.) \pm 0.08 \:(syst.) \:\text{fm}^2
\end{equation*}
in the \Q2 region of
$  0.03 \le Q^2 \le 0.16 \; \text{GeV}^2/c^2 $ (7,800 events)\cite{ivo}.
This result is well inside
the limits determined by the WA89 collaboration \cite{wa89_radius_paper},
$0.4~\fm2 \le \r2_{\Sigma^-} \le 1.4~\fm2$ (Fig.~\ref{fig:sigmaresult}).

For the negative pion, a monopole electric form factor is used. We find
\begin{equation*}
  \label{equ:pion_selex}
  \r2_{\pi} = 0.45 \pm 0.03 \:(stat.) \pm 0.07 \:(syst.) \:\fm2,
\end{equation*}
where
$  0.03 \le Q^2 \le 0.20 \; \text{GeV}^2/c^2.$ (12,000 events)\cite{kvo_phd}.
This result is in excellent agreement with
the so far best direct measurement \cite{amendolia86pion} of
{\nobreak $\r2_{\pi} = 0.44 \pm 0.01 \:\fm2$} as well as a recent
calculation which takes into account form factor measurements in both
space-like and time-like regions \cite{geshkenbein98}:
{\nobreak $\r2_{\pi} = 0.463 \pm 0.005 \:\fm2.$}

Major contributions to the systematic error come from the \Q2 resolution,
uncertainties in the corrections for trigger efficiency, and beam
contamination by other particles, particularly \PgXm. Significant improvement
is expected for all of these when advanced reconstruction and simulation
software is used to refine the data sample. \Q2 will be determined from
all kinematic variables and events with identified \PgSm\ decays accepted
as well. We anticipate a statistical error
of less than 10~\% in the final analysis of the \PgSm\ radius.

\section{Summary}
\label{sec:concl}

A measurement of the \PgSm\ and \Pgpm\ mean squared charge radii has
been performed by elastic hadron-electron scattering.
A preliminary analysis yields a \PgSm\ radius which is within 2$\sigma$
of the proton radius. The \Pgpm\ radius is in excellent agreement with 
previous experiments.

%
%



\end{document}